# Comment on "Thermodynamic properties of rock-salt ZnO" by J. Leitner, M. Kamrádek and D. Sedmidubský
# [ Thermochimica Acta 572 (2013) 1-5 ]

P.S. Sokolov,[1]  O.O. Kurakevych,[2]  A.N. Baranov,[3]  and  V.L. Solozhenko[1*]

[1] *LSPM–CNRS, Université Paris Nord, 93430 Villetaneuse, France*

[2] *IMPMC, Université P&M Curie, 75005 Paris, France*

[3] *Moscow State University, Moscow 119991, Russia*

Thermodynamic properties of rock-salt zinc oxide (*rs*-ZnO) are interesting from both fundamental (phase diagram [1,2], transition hysteresis as a function of time scale and nanosize [3], etc.) and applied (new advanced materials [4]) points of view. Very recently *Leitner et al.* [5] have tried to extract the thermodynamic data of rock-salt ZnO from *ab initio* and experimental data available in the literature. However, neglecting the strongly pronounced kinetic features of the pressure-induced phase transition in ZnO below 1000 K [3] rendered these results ambiguous. Moreover, they contradict the experimental data (*in situ* observations at high pressure and high temperature (HPHT) [1,2], and calorimetric measurements at ambient pressure [6]). In this Comment we would like to clarify the situation and discuss basic experimental data for the construction of a full set of thermodynamic properties of *rs*-ZnO.

In the work of *Leitner et al.* [5], the key thermodynamic value, $\Delta_{tr}G^0 = 23.12$ kJ mol$^{-1}$, corresponding to the wurtzite–to–rock-salt phase transition in ZnO at 0.1 MPa and 298.15 K, has been estimated exclusively from one of the most questionable values, the room-temperature equilibrium transition pressure, $P_{tr}$. The choice of $P_{tr} = 9.6$ GPa based on averaging of *ab initio* estimations (that vary from 7.4 GPa [7] to 12.7 GPa [8]) and one randomly chosen value of the onset pressure of kinetically hindered phase transition at 300 K (that also varies from 7.5 GPa [9] to 10.0 GPa [10]). All this leads to completely wrong equilibrium *P-T* line in the HPHT region, where the transition is governed by thermodynamics, and equilibrium experimental data are available [1,2].  However, this fact was completely neglected by *Leitner et al.* [5].

Single-phase *rs*-ZnO has been recently recovered at ambient conditions in the form of a nanocrystalline bulk [4], and the enthalpy of the *w*-ZnO-to-*rs*-ZnO phase transition, $\Delta_{tr}H = 10.2 \pm 0.5$ kJ mol$^{-1}$ (at 400 K) was evaluated using differential scanning calorimetry [6]. This experimental value is more than 2 times lower than the "recommended" one, $\Delta_{tr}H^0$(298.15 K) = 23.93± 3.11 kJ mol$^{-1}$ claimed by *Leitner et al.* [5] under suggestion that $P_{tr}$ = 9.6 GPa. It should be noted that with more reasonable choice of the 300-K transition pressure i.e. $P_{tr} \approx 5.8$ GPa [2], the $\Delta_{tr}G^0 = \int_{P0}^{Ptr}(V_m^w - V_m^{rs})dp$ value makes ~14 kJ mol$^{-1}$, and $\Delta_{tr}H^0 \approx 15$ kJ mol$^{-1}$, in agreement with the experimental $\Delta_{tr}H^0$ value at 400 K [6].

---

[*] Corresponding author.  E-mail: vladimir.solozhenko@univ-paris13.fr

Thus, all thermodynamic data of *rs*-ZnO reported by *Leitner et al.* [5] should be reconsidered in order to make them compatible with experimental equilibrium *P-T* data on ZnO phase transitions at HPHT, as well as with results of calorimetric measurements.

**References:**

1. K. Kusaba, Y. Syono, T. Kikegawa, Phase transition of ZnO under high pressure and temperature, Proc. Jap. Acad. B 75 (1999) 1-6.

2. F. Decremps, J. Zhang, R.C. Liebermann, New phase boundary and high-pressure thermoelasticity of ZnO, Europhys. Lett. 51 (2000) 268-274.

3. V.L. Solozhenko, O.O. Kurakevych, P.S. Sokolov, A.N. Baranov, Kinetics of the wurtzite-to-rock-salt phase transformation in ZnO at high pressure, J. Phys. Chem. A 115 (2011) 4354–4358.

4. A.N. Baranov, P.S. Sokolov, V.A. Tafeenko, C. Lathe, Y.V. Zubavichus, A.A. Veligzhanin, M.V. Chukichev, V.L. Solozhenko, Nanocrystallinity as a route to metastable phases: rock salt ZnO, Chem. Mater. 25 (2013) 1775-1782.

5. J. Leitner, M. Kamrádek, D. Sedmidubský, Thermodynamic properties of rock-salt ZnO, Thermochim. Acta 572 (2013) 1-5.

6. P.S. Sokolov, A.N. Baranov, Zh.V. Dobrokhotova, V.L. Solozhenko, Synthesis and thermal stability of rock-salt ZnO in salt nanocomposites, Russ. Chem. Bull. 59 (2010) 325-328.

7. A. Seko, F. Oba, A. Kuwabara, I. Tanaka, Pressure-induced phase transition in ZnO and ZnO-MgO pseudobinary system: A first-principles lattice dynamics study, Phys. Rev. B 72 (2005) 024107.

8. M. Kalay, H.H. Kart, S Özdemir Kart, T, Çağin, Elastic properties and pressure induced transition of ZnO polymorphs from first-principal calculations, J. Alloys Compd. 484 (2009) 306-310.

9. F. Decremps, J. Zhang, B. Li, R.C. Liebermann, Pressure-induced softening of shear modes in ZnO, Phys. Rev. B 63 (2001) 224105.

10. L. Gerward, J. S. Olsen, The high-pressure phase of zincite, J. Synchrotron Rad. 2 (1995) 233-235.